\documentclass[11pt]{article}
\usepackage{moriond,epsfig}
\usepackage[latin1]{inputenc}
\usepackage[OT1]{fontenc}

\def\Journal#1#2#3#4{{#1} {\bf #2}, #3 (#4)}

\def\be{\begin{equation}}
\def\ee{\end{equation}}
\def\bea{\begin{eqnarray}}
\def\eea{\end{eqnarray}}

\def\chisq{$\chi$$^2$}
\def\loglik{log(likelihood)}

\def\sext{$\sigma$$_{\rm ext}$}
\def\c2nor{\chi^2}

\def\epo{$E_{\rm p}$}
\def\epi{$E_{\rm p,i}$}
\def\eiso{$E_{\rm iso}$}
\def\tb{$t_{\rm b}$}
\def\lpiso{$L_{\rm p,iso}$}
\def\liso{$L_{\rm iso}$}
\def\ega{$E_{\gamma}$}
\def\epeiso{$E_{\rm p,i}$ -- $E_{\rm iso}$}
\def\epega{$E_{\rm p,i}$ -- $E_{\gamma}$}
\def\nufnu{$\nu$F$_\nu$}
\def\epeisotb{$E_{\rm p,i}$ -- $E_{\rm iso}$ -- $t_b$}
\def\sax{{\it Beppo}SAX}
\def\swift{{\it Swift}}
\def\fermi{{\it Fermi}}
\def\konus{{\it Konus}/WIND}

\def\omegam{$\Omega$$_{\rm M}$}
\def\omegal{$\Omega$$_{\Lambda}$}
\def\h0{H$_{\rm 0}$~}

\begin{document}
\vspace*{4cm}
\title{COSMOLOGY WITH GAMMA--RAY BURSTS}

\author{ L. AMATI }

\address{INAF - IASF Bologna, via P. Gobetti 101,\\
I-40129 Bologna, Italy}

\maketitle\abstracts{Gamma--Ray Bursts (GRBs) are the brightest sources in the 
universe, emit mostly in the hard X--ray energy band and have been detected at 
redshifts up to $\sim$8.1. Thus, they are in principle very powerful probes for 
cosmology. I shortly review the researches aimed to 
use GRBs for the measurement of cosmological parameters, which are mainly based on 
the correlation between spectral peak photon energy and total radiated energy or 
luminosity. In particular, based on an enriched sample of 95 GRBs, I will provide 
an update of the analysis by Amati et al. (2008) aimed at extracting information on 
\omegam{} and, to a less extent, on \omegal, from the \epeiso{} correlation. I also 
briefly discuss the perspectives of using GRBs as cosmological beacons for high 
resolution absorption spectroscopy of the IGM (e.g., WHIM), and as tracers of the 
SFR, up to the "dark ages" ($z > 6$) of the universe.
} 


\section{Introduction}

After $\sim$40 years since its discovery, the GRB phenomenon is still one of the 
most intriguing and hot topics in modern astrophysics. Indeed, despite the huge 
observational advances occurred since the late 90s, with the 
discovery of the afterglow emission, optical counterparts, host galaxies, the 
determination of the cosmological distance scale and huge luminosity and the 
evidences of association with peculiar SNe, our understanding of GRBs origin and 
physics is still affected by several open issues \cite{Meszaros06}. Among these, one of the most 
intriguing and debated is the possible use of GRBs as comological probes, 
which has been proposed in the last few years by several authors, following the 
mounting evidence that they are the brightest and farthest sources in the 
universe. In particular, many efforts have been done in 
order to extract information on cosmological parameters in an independent, or 
complementary, way to type Ia SNe and other cosmological probes (e.g., BAO, galaxy 
clusters, the CMB) by "standardizing" GRBs with the so called spectrum--energy 
correlations. Also, the high X--ray flux and the association of long GRBs with the 
death of young massive stars prompted the investigation of GRBs as background 
sources for high resolution spectroscopy of the IGM with next generation 
experiments and as tracers of the star formation rate (SFR) up to the re--ionization 
epoch.

In this article, after summarizing the properties that make GRBs potentially 
powerful cosmological probes (Section 2), I will discuss and update the analysis 
aimed at estimating cosmological parameters by using the \epeiso{} correlation, the 
simplest and first discovered among spectrum--energy correlations (Sections 3 and 
4). Then, I will review (Section 5) the results on cosmological parameters 
obtained by using other spectrum--energy correlations found by 
adding to \epi{} and \eiso{} 
a third observable. Historically, these correlations were the first to be 
used to this purpouse since 2004. Methods based on the joint use of 
spectrum--energy correlations with Type Ia SNe or other GRB correlations are 
outlined in Section 6. Finally, in Section 7 I briefly discuss the possible use 
of GRBs as cosmological beacons and tracers of the star formation history of the 
universe.

For reasons of space, the citations in the text cannot be exhaustive, and the
given references are reviews or examples. 
The analysis reported in Sections 2, 3 and 4 are 
based on data available as of April 2009 and have been performed specifically for 
this work.

\begin{figure*}
\centerline{\psfig{figure=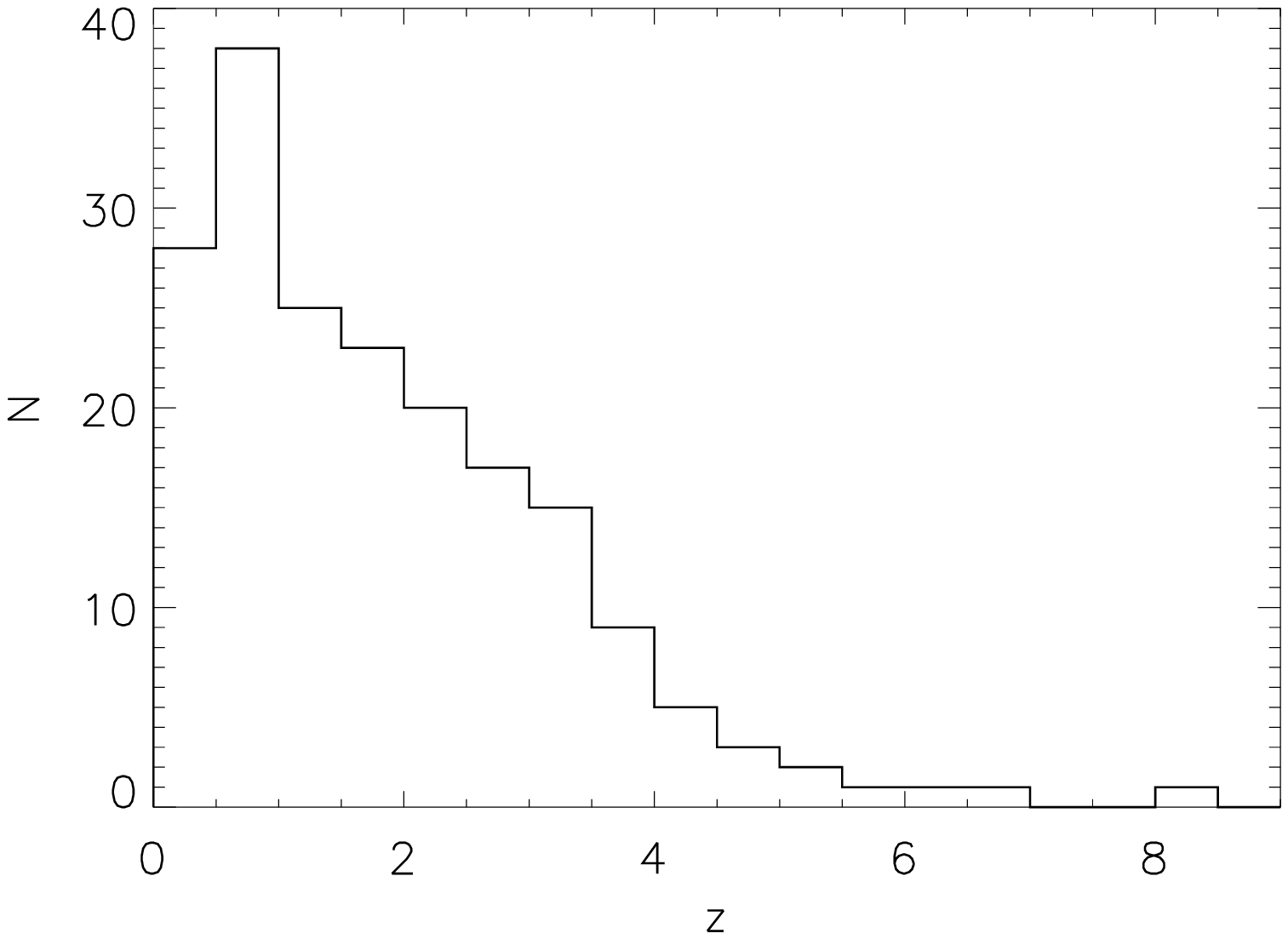,width=8.5cm}\psfig{figure=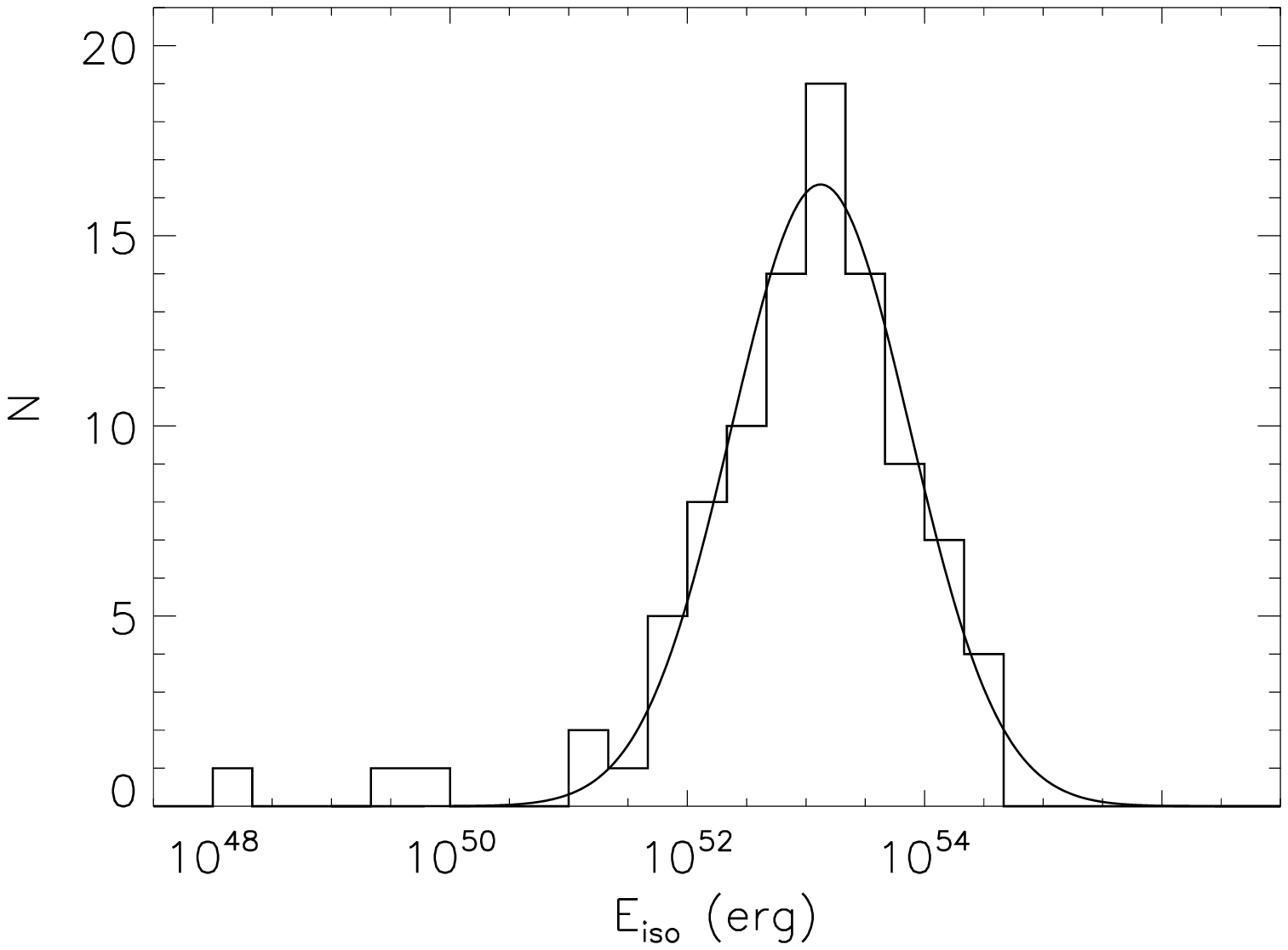,width=8.5cm}}
\caption{Left: redshift distribution for the 189 GRBs with known redshift as of 
April 2009. Right: \eiso{} distribution for the 95 GRBs with known redshift and 
spectral parameters as of April 2009.
}
\end{figure*}

\section{Gamma--Ray Bursts as cosmological probes}

In the last years, the use of Type Ia SNe as standard candles, combined with CMB 
measurements, has revolutioned our view of the history of the cosmic expansion of the 
universe. Indeed, within the standard CDM cosmological model the evidence, based on 
CMB observations and the implications of inflation, that the universe is flat 
($\Omega$=1) and the location of high--z SNe Ia in the Hubble diagram imply that 
the 
universe is presently accelerating and that $\sim$73\% of $\Omega$ is determined by 
an unknown and mostly unpredicted component or field (e.g., dark energy, 
quintessence, cosmological constant) \cite{Perlmutter98,Riess98,Riess04}. 
However, SN Ia as standard candles are affected 
by possible systematics, like, e.g., different explosion mechanisms and progenitor 
systems, evolution with $z$, possible dependence on $z$ of the light curve shape 
correction for luminosity normalization, signatures of evolution in the colours, 
correction for dust extinction, anomalous luminosity--color relation, contaminations 
of the Hubble Diagram by no--standard SNe-Ia and/or bright SNe-Ibc (e.g. HNe)
\cite{Dellavalle08}. In 
addition, this sources are found only up to moderate redshift ($\sim$1.4--1.7).

Thus, the quest for alternative astrophysical sources capable to provide estimates of 
the cosmological parameters in an independent way and at higher $z$ with respect to 
SNe Ia is a central topic in modern astrophysics. The sources under 
investigation for this purpouse include, e.g., galaxy 
clusters and BAO, but a lot of interest has been raised in the last years by the 
redshift and luminosity properties of GRBs.
 
In Figure~1, I show the updated distributions of $z$ (189 events with measured 
redshift) and \eiso{} (95 events with measured redshift and spectral parameters) of 
GRBs as of April 2009. The redshift values were taken from the GRB table by J. 
Greiner \footnote{http://www.mpe.mpg.de/jcg/grbgen.html} and references therein, 
whereas the values of \eiso{} were computed based on the spectral parameters and 
fluences reported in Amati et al. (2008) \cite{Amati08}, 70 events, and Amati et al. 
(2009) \cite{Amati09}, 25 more events, 
and by assuming a standard $\Lambda$CDM cosmology with \h0 = 70 km s$^{-1}$ 
Mpc$^{-1}$, \omegam{} = 0.27 and \omegal{} = 0.73. As can be seen, 
GRBs are the brightest 
sources in the universe, with values of the isotropic--equivalent radiated energies, 
\eiso, that can exceed 10$^{54}$ erg, emit most of their radiated energy in the hard 
X--rays, and thus are not affected by dust extinction problems which affect, e.g., 
type Ia SNe, and show a redshift distribution extending at least up to $\sim$8.1, 
much above that of any other class of astrophysical sources. 

Thus, in principle GRBs are the most suitable cosmological probes. However, as can 
be seen in Figure~1, they are not standard candles, showing radiated energies, and 
luminosities, spanning several orders of magnitude. In the past it was proposed 
that the collimation--corrected radiated energy, \ega\ (see Section 5) could be 
clustered at around $\sim$10$^{51}$ erg \cite{Frail01,Berger03}, but this evidence was not confirmed by 
subsequent observations. The investigation of GRBs as a new and alternative tool 
for the measurement of cosmological parameters was then prompted by the discovery of a 
strong correlation between the spectral peak photon energy, a quantity independent 
on the cosmological model, and the event intensity (radiated energy, average luminosity, peak luminosity), which depends on the assumed cosmological parameters.
This correlation and the methods proposed to derive from it information on 
cosmological parameters are the subject of the next three Sections.

\begin{figure*}
\centerline{\psfig{figure=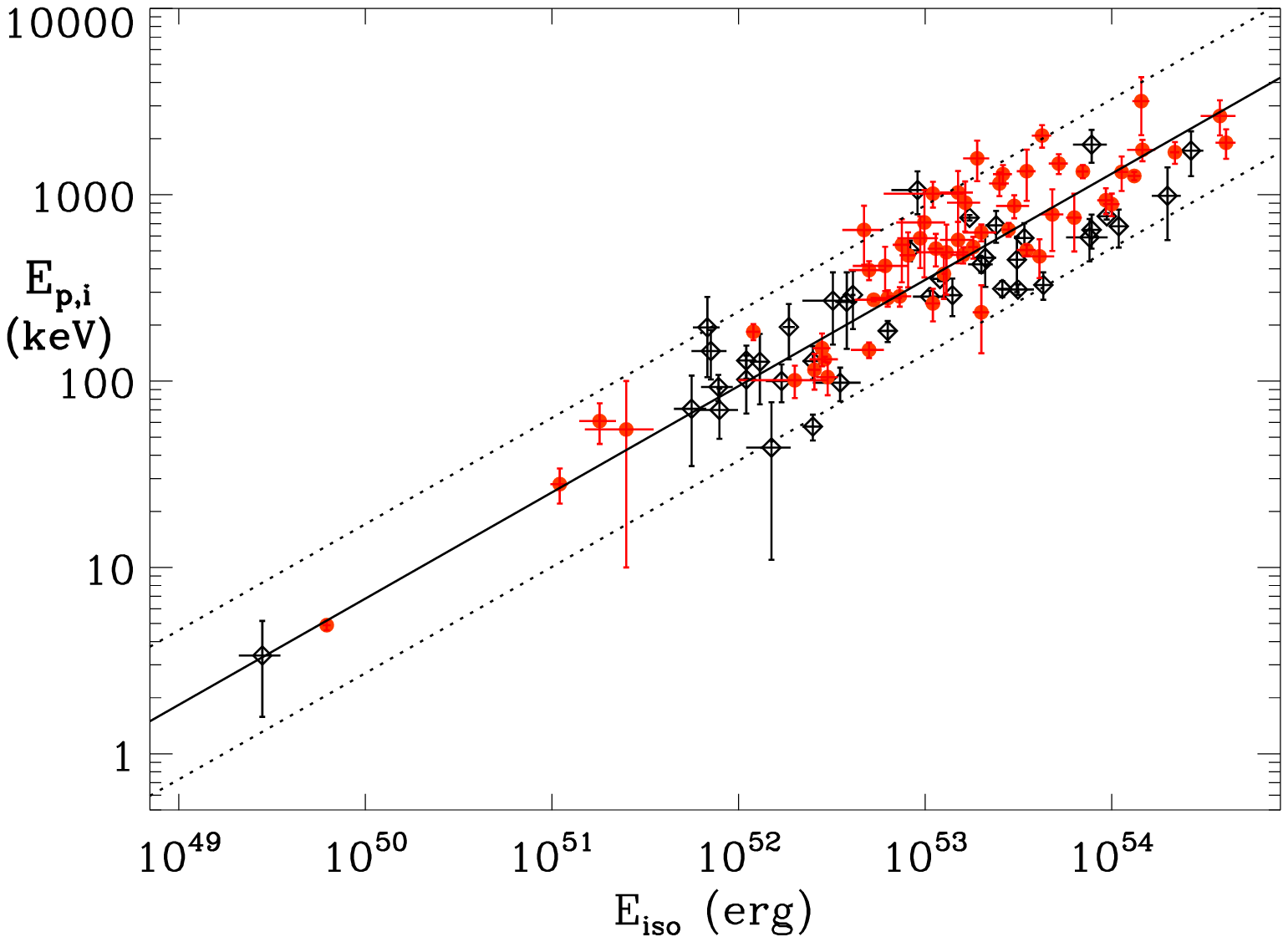,width=8.5cm,height=6cm}\psfig{figure=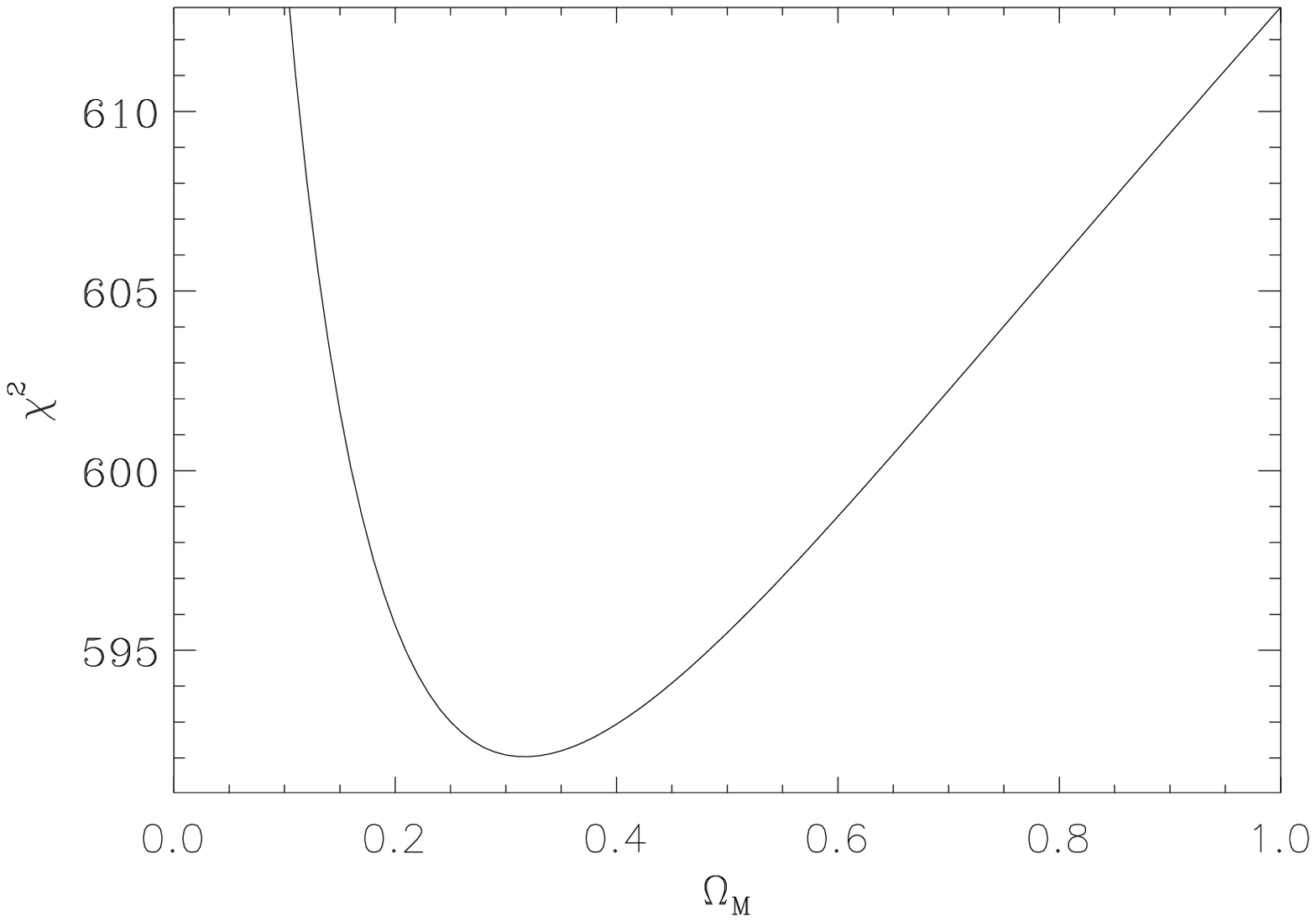,width=8.5cm,
height=6cm}}
\caption{Left: \epeiso{} correlation withe the prsent sample of 95 GRBs with known
redshift and \epi. Red dots are those GRBs localized by \swift. 
Right: \chisq{} value of the fit of the updated \epeiso{} correlation (95 GRBs) 
with a simple power--law as a function of the
value of \omegam{} assumed to compute the \eiso{} values. A flat univese is assumed.
} 
\end{figure*}

\section{The \epeiso{} correlation}

GRB spectra are non thermal and are well described by a smoothed broken power--law 
("Band" function) with low and high energy photon indices in the ranges 
$\sim$0.5--1.5 and $\sim$2.1--3.5, respectively \cite{Band93,Kaneko06}. Thus, 
when expressed in terms of 
\nufnu, GRB spectra show a peak. The photon energy at which this peak occurs is 
hence called "peak energy" and indicated as \epo{} when referring to the observed 
spectrum or \epi{} for the cosmological rest--frame (i.e., "intrinsic") spectrum. 
\epi{} values range from a few keV up to several thousends of keV and its  
distribution has the shape of a Gaussian centered at around 200--300 keV with a low 
energy tail \cite{Amati06}. This spectral parameter is a relevant observable for models of the 
physics of GRB prompt emission \cite{Zhang02}, whose understanding is one of the 
main still open issues in this field of research.

Evidence for a strong correlation between \epi, and \eiso{} was first reported by 
Amati et al. (2002)\cite{Amati02}, based on a limited sample of \sax{} GRBs with 
known redshift. This correlation was later confirmed and extended to softer/weaker 
events (X--Ray Flashes, XRFs) by measurements by other satellites, mainly HETE--2, 
\konus{} and, more recently also \swift{} 
and \fermi/GBM \cite{Amati06,Amati08,Ghirlanda08}
The recent estimates of $z$ for some short GRBs provided the 
evidence that the \epeiso{} correlation holds only for long GRBs \cite{Amati06b,Ghirlanda09}, with the exception 
of the peculiar sub--energetic GRB\,980425. It was also found that the correlation 
holds as well if \eiso{} is substituted with the average or peak luminosity (\liso{} 
and \lpiso, respectively \cite{Lamb04,Yonetoku04}), which is 
not surprising given that these "intensity indicators" are strongly correlated. In 
Figure~2, I show the \epeiso{} correlation for the most updated (April 2009) sample 
of GRBs with known $z$ and \epi. The main features of the \epeiso{} correlation are 
that it extends over several orders of magnitude both in \epi{} and \eiso, it can be 
modeled by a power--law with slope $\sim$0.5 and it is characterized by an 
extra--scatter, with respect to Poissonian fluctuations, of $\sim$0.2 dex 
\cite{Ghirlanda08,Amati06,Amati08}.

As already discussed by several 
authors \cite{Zhang02,Lamb05,Amati06}, this observational evidence has relevant 
implications for the geometry and physics of GRB prompt emission and can be used to 
identify and understand sub--classes of GRBs (e.g., short, sub--energetic, XRFs). 
In the recent years some authors argued that the correlation may be an artifact of, 
or at least significantly biased by, a combination of selection effects due to 
detectors sensitivity and energy thresholds \cite{Band05,Nakar05,Butler09}. 
However, the fact that GRBs detected, 
localized and spectroscopically characterized by different instruments all follow 
the same \epeiso{} correlation, as shown by Amati et al. (2009) \cite{Amati09}
and can also be seen in Figure~2 by comparing the location 
of \swift{} GRBs with respect to those detected by other instruments, supports the 
hypothesis of a low impact of selection and detectors threshold effects. Moreover, 
time resolved analysis of large samples of GRBs provide evidence that the 
correlation holds also within single bursts \cite{Liang04,Firmani08}, thus pointing to a physical origin of 
it.

\begin{figure*}
\centerline{\psfig{figure=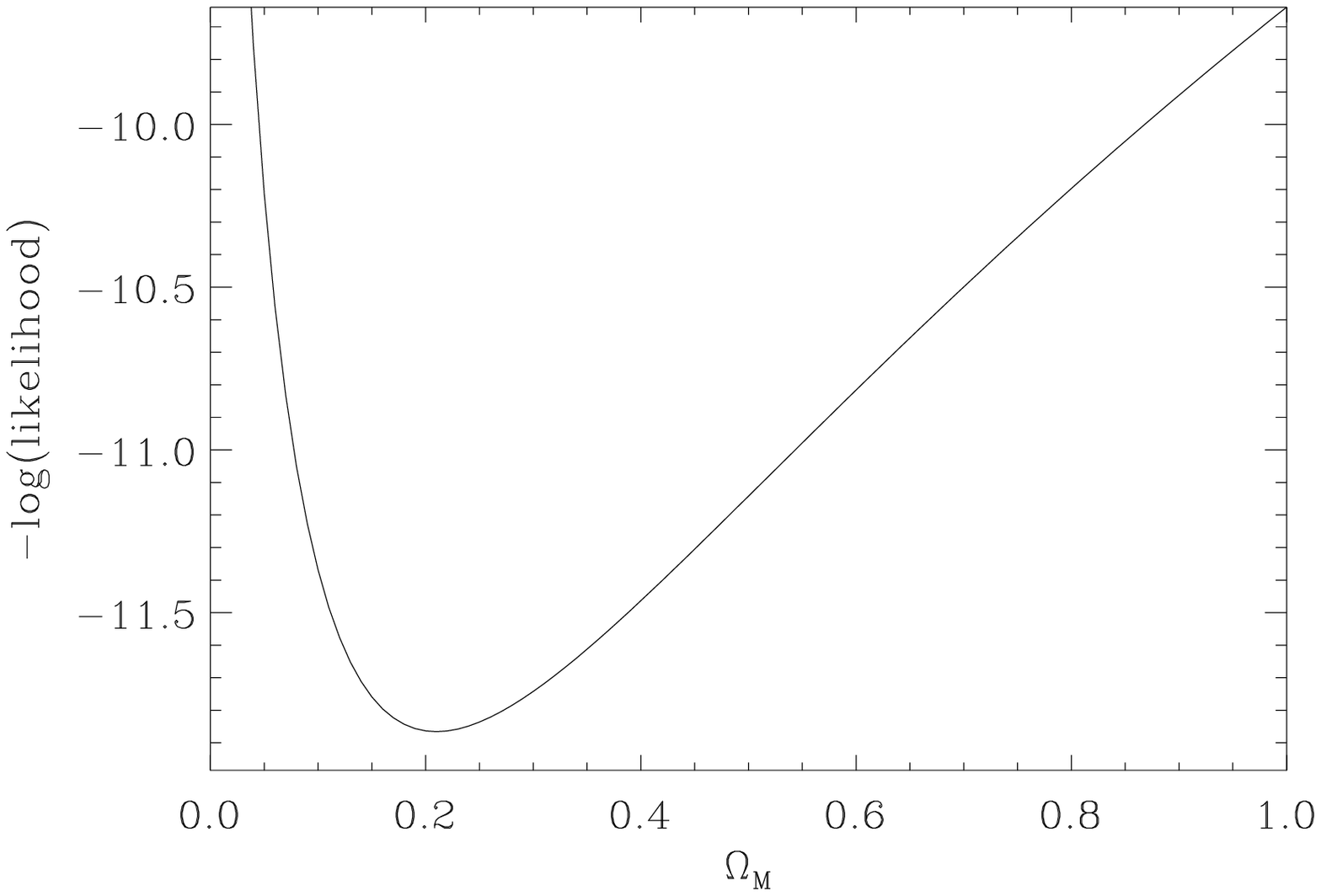,width=8.5cm,height=6cm}\psfig{figure=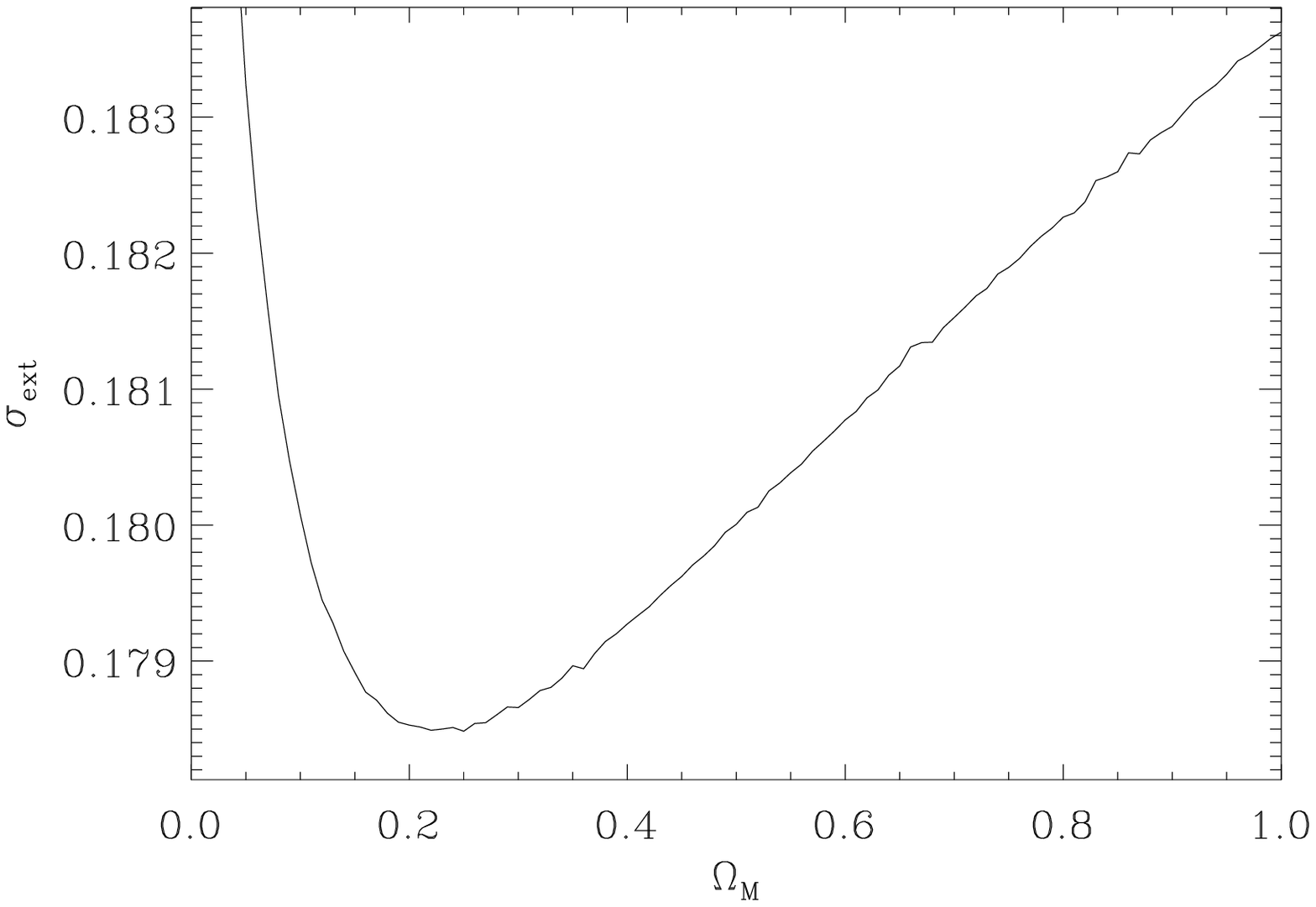,width=8.5cm,
height=6cm}}
\caption{
Values of \loglik (left) and \sext{} (right) of the fit of the updated \epeiso{} 
correlation (95 GRBs) 
with a maximum likelihood method accounting for extrinsic variance (see text)
as a function of the
value of \omegam{} assumed to compute the \eiso{} values. A flat univese is assumed.
} 
\end{figure*}

\begin{figure*}
\centerline{\psfig{figure=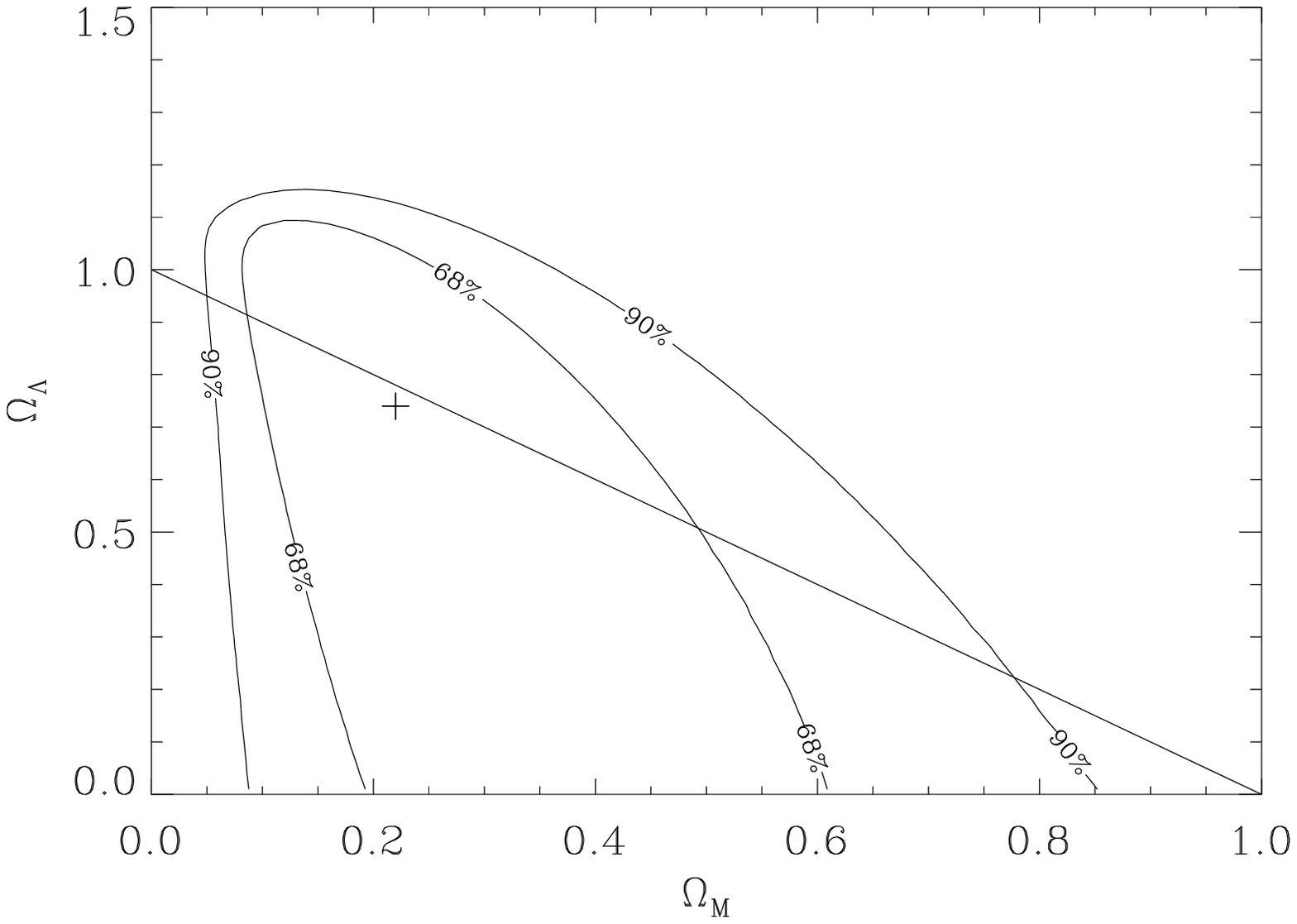,width=8.5cm,height=6cm}\psfig{figure=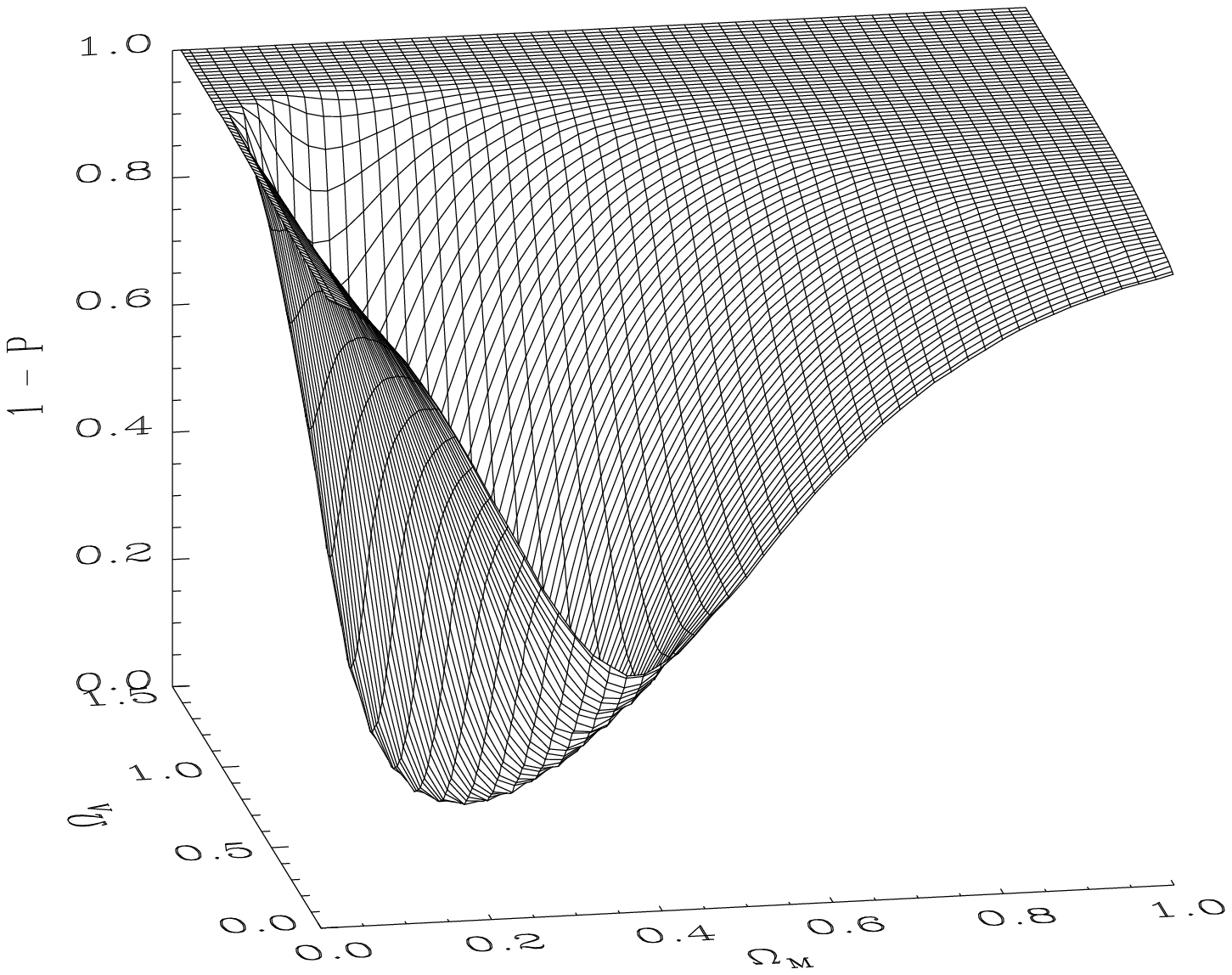,width=8.5cm,
height=6cm}}
\caption{
Contour (left) and surface (right) plots showing the probability associated to 
\omegam{} and \omegal{} found by fitting the updated \epeiso{}
correlation (95 GRBs)
with a maximum likelihood method accounting for extrinsic variance (see text) and
releasing the hypothesis of a flat universe. The cross in the left panel indicates 
the best fit values.
} 
\end{figure*}

\section{Estimating cosmological parameters with the \epeiso{} correlation}

As discussed in the previous Section, the \epeiso{} correlation is highly 
significant, holds for all long GRBs with known redshift and \epi{} and is likely not 
strongly affected by selection and detectors threshold effects. Thus, given that it 
links a cosmology independent quantity, \epi, to the burst radiated energy or 
luminosity, in principle it could be used to "standardize" GRBs, in a way similar 
to what is done with SNe Ia with the "Phillips" relation. However, the dispersion 
of the points around the best fit power--law is significanlty in excess to the 
Poissonian one, indicating the presence of an extrinsic variance of unknown origin. 
In addition, given the lack of a sufficient number of GRBs at very low  or at the
same redshift (Figure~1), the 
correlation cannot be calibrated, as can be done, instead, for Type Ia SNe. Because 
of this problems, in the last years the "cosmological use" of this correlation, 
and/or the \epi{}--\lpiso{} correlation, consisted in the estimate of pseudo--redshifts 
for those GRBs without measured redshift. This can be done by simply studying the 
track of a GRB in the \epeiso{} plane as a function of redshift, or by using quantities involved in the 
correlation to build a pseudo--redshift estimator \cite{Atteia03}. Some authors 
applied these methods on large sample of GRBs in order to reconstruct the 
luminosity function or, assuming the association of GRB with very massive stars, 
the star formation rate (SFR) evolution (Section~7).

However, recently Amati et al. (2008) \cite{Amati08} have shown that the \epeiso{} 
correlation can also be used to obtain information on cosmological parameters. 
Their work was prompted by the evidence that, in the assumption of a flat universe, 
the trend of the \chisq{} of the fit with a simple powerlaw as a function of the 
value of \omegam{} adopted to compute the luminosity distance and hence the values of 
\eiso{} shows a nice parabolic shape minimizing at \omegam$\sim$0.3, as can be seen 
in Figure~2. This is a qualitative but simple, and independent on other 
cosmological probes, indication that if the universe is flat, as predicted by 
inflation and implied by CMB measurements, the universe expansion is presently 
accelerating and an unknown component or field (e.g., dark energy, quintessence, 
cosmological constant) is dominating over matter and/or gravity. In order to 
quantify the estimate of \omegam, Amati et al. (2008) adopted a likelihood method 
which accounts for uncertainties on both X and Y quantities and parametrizes the 
extrinsic variance (i.e. the variance in excess to the Poissonian one) of the data, 
\sext{}. In this way, they found \omegam{} = 
0.15$_{-0.11}^{+0.25}$ at 68\% c.l. and \omegam{} $<$1 at a 
significance level higher than 99\%. This result is fully consistent with that 
obtained with type Ia SNe. By means of simulations, they also showed that 
with the substantial increase of the number of GRBs with known $z$ and \epi{} 
expected in the next years, these constraints will be significantly reduced. 

In Figures 3 I show the results obtained by repeating the same analysis on 
the updated sample of 95 GRBs. As can be seen, both the $-$log--likelihood and 
\sext{} minimize for \omegam $\sim$0.2. In particular, I find \omegam{} = 
0.21$_{-0.13}^{+0.27}$ at 68\% c.l. and \omegam{} = 0.21$_{-0.16}^{+0.53}$ at 90\% 
c.l. These constraints are slightly tighter than those obtained by Amati et al. 
(2008), confirming the expected effect of the sample enlargement. Moreover, as 
can be seen in Figure~4, by
releasing the assumption of a flat universe, the best--fit values of \omegam{} and 
\omegal{} are 0.22 and 0.74, respectively, i.e. very close to the standard cosmology 
values and to the flat universe hypothesis. Also in this case, even if at 68\% c.l. 
they still provide only an upper 
limit to \omegal, the contour confidence levels are tighter than what found by 
Amati et al. (2008).

\section{Cosmology with three-parameters spectrum--energy correlations}

Soon after the first detections of GRB optical counterparts, it was found that in 
some cases the optical afterglow light curve shows a steepening of its power--law 
decay \cite{Harrison99,Ghirlanda04}. Within the standard fireball -- external shock scenario for the afterglow 
emission, this "break" can be interpreted as due to collimated emission 
\cite{Sari99} (even 
though other explanations are possible). In this view, the jet opening angle can be 
derived from the break time \tb{} by making some assumptions on the circum--burst 
medium average density and profile and on the efficiency of conversion of the 
fireball kinetic energy into radiated energy. The jet opening angle, in turn, can 
be used to derive the collimation--corrected, or "true", radiated energy, \ega,  
from \eiso. As mentioned in Section~2, \ega{} is sitll not standard and 
is tipically in the range from $\sim$5$\times$10$^{49}$--10$^{52}$ erg.

In 2004 it was found that when substituing \eiso{} with \ega the \epeiso{} correlation 
becomes tighter, i.e. its extrinsic scatter reduces by a factor of $\sim$2
\cite{Ghirlanda04}. Even if 
based on a rather low number of events, this evidence prompted the first systematic 
investigations of GRBs as cosmological rulers \cite{Dai04,Ghirlanda04,Ghirlanda06}. 
Despite the 
advantage of a reduced scatter with respect to the \epeiso{} correlation, the problem 
of the lack of calibration with low $z$ events cannot be solved anyway, and 
different methods were proposed in order to avoid "circularity". The most common 
are the so called "scatter methods", consisting in fitting the correlation for each 
set of cosmological parameters under study, deriving a \chisq{} distribution and use 
it to obtain best fit values and confidence intervals. This can be done either 
directly in the \epega{} plane or in the Hubble diagram obtained by deriving \ega{} 
from \epi{} and hence the luminosity distance from \ega{} and the measured fluence. 
More sophisticated methods based on Bayesian statistics were also proposed
\cite{Ghirlanda06}. 

The constraints on \omegam{} and the limits to \omegal{} obtained with the \epega{} 
correlation were similar to those derived a few years later from the \epeiso{} 
correlation and described in the previous Section. The main drawbacks that 
prevented, up to now, the expected improvements in the accuracy and 
reliability of the cosmological parameters estimates with this method include: i) 
the very slow increase of GRBs with evidence of a break in the optical afterglow 
light curve, mainly due to the lack of systematic monitoring (the number of GRBs 
that can be used for the \epega{} correlation are $\sim$25\% with respect to the 
\epeiso{} correlation); ii) the evidence from \swift/XRT measurements of the X--ray 
afterglow that, contrary to what expected in the basic jet scenario, in several 
cases there are not X--ray breaks or the breaks are achromatic; iii) the debate on 
the real dispersion and possible existence of outliers of the \epega{} correlation
\cite{Campana07,Ghirlanda07}; 
iv) the fact that the \epega{} correlation is model dependent, i.e. requires 
assumptions on the circum--burst density profile and, more in general, a jet model. 
Concerning points ii) and iv), it was noted that the correlation between \epi, 
\eiso{} and \tb{} holds even without the need of a jet interpretation, i.e. at a purely 
empirical level \cite{Liang05}. Thus, also the \epeisotb{} correlation was investigated for the 
estimate of cosmological parameters. However, under this respect it is still 
affected by the low number of 
events that can be used, the existence of possible outliers and the uncertainty on 
its true dispersion.

In 2006, it was also found that the dispersion of the \epi{}--\lpiso{} correlation 
decreases substantially when including the "high signal time scale" T$_{0.45}$, a 
parameter often used in GRB variability studies. Thus, also this correlation was 
proposed as a tool to standardize GRBs, similarly to the \epega{} and \epeisotb{} 
correlations, but with the advantage of a higher number of events, being based on 
prompt emission properties only. However, subsequent analysis on larger samples 
showed that the extrinsic scatter of this correlation may not be significantly lower 
than that of the simple \epeiso{} or \epi{}--\lpiso{} 
correlations \cite{Rossi08,Collazzi08}.

\section{Calibrating GRBs with SNe Ia and multi--correlation studies}

As mentioned in the previous Sections, one of the most liming features of 
spectrum--energy correlations as tools to standardize GRBs is the lack of low 
redshift GRBs, or of a sufficient number of GRBs at the same redshift, allowing to 
calibrate them. On the other hand, if one believes that SN Ia are reliable distance 
indicators, then can use them to calibrate GRB spectrum--energy correlations and 
take advantage of the GRB redshift distribution in order to extend the Hubble 
diagram from z $\sim$ 1.7 up to $\sim$8. This approach has been followed by several 
authors \cite{Kodama08,Liang08}, 
allowing them not only to tighten the constraints on \omegam{} and \omegal{} 
but also to obtain information on the dark energy equation of state and its 
evolution, or to test cosmological models alternative to the standard $\Lambda$CDM.

The obvious drawback of this use of GRBs for cosmology is that it introduces a 
"circularity" with type Ia SNe, i.e., GRBs are no more independent probes and all 
the systematics and uncertainties associated with SNe propagates into the
results obtained with this method.

The spectrum--energy correlations discussed in previous Sections are the tightest 
but not the only ones linking GRB observables to their luminosity. For instance, 
significant correlations were found between prompt emission variability and peak 
luminosity or between prompt emission time--lag and luminosity. Some authors 
developed methods for putting together several correlations in order to derive 
estimates of cosmological parameters \cite{Schaefer07}. 
However, adding to spectrum--energy 
correlations more dispersed correlations adds more uncertainties, 
thus preventing a significant 
improvement with respect to using spectrum--energy correlations alone.

\section{Gamma--Ray Bursts as cosmological beacons and SFR tracers}

Besides the estimate of cosmological parameters, GRBs are also very promising tools 
for cosmology under other respects. The association of long GRBs with peculiar type 
Ib/c SNe or hypernovae, and thus the death of very massive stars, is supported both 
by theories and observations \cite{Woosley06}. Thus, given their huge luminosity and redshift 
distribution extending up to at least $z$ $\sim$ 8, GRBs may be considered powerful 
and unique tracers of the SFR evolution up to the re-ionization epoch. For instance, 
the recent detection of GRB\,090423 at $z$ $\sim$ 8.1 is a simple and direct evidence that 
stars were already there at about 600 millions of year from the big--bang and with 
explosion mechanism not markedly different from that of stars born several billions 
of years later \cite{Salvaterra09}. Several authors addressed this issues, either by comparing 
directly the GRB redshift distribution with the SFR up to $z$ $\sim$ 4 reconstructed 
from other observations, or by reconstructing the GRB luminosity function and its 
evolution by computing the pseudo--redshift of large numbers of GRB based on 
spectrum--energy correlations \cite{Yonetoku04}. The results of these analysis indicate that GRBs 
are a biased tracer of the SFR evolution, which may be due to the fact, supported 
both by theory and observations, that GRBs are produced by low metallicity stars in 
low metallicity galaxies. Under this respect, GRBs provide information on the 
metallicity evolution \cite{Li08}.

Another interesting and promising cosmological use of GRBs is to use their X--ray 
afterglow emission as background source for X--ray high resolution spectroscopy of 
the inter--galactic medium (IGM) and of the host galaxies inter--stellar medium 
(ISM). This kind of investigations is the subject of future missions under study, 
like, e.g., the EDGE mission proposed to the ESA Cosmic Vision \cite{Piro09}
or the XENIA 
mission submitted to the NASA Decadal Survey. As discussed, e.g., by Branchini et 
al. (2009) \cite{Branchini09}, with state of the art X--ray microcalorimeters, 
allowing energy resolutions of the order of $\sim$2--3 eV in the 0.2--2 keV energy 
range, an effective area of $\sim$1000 cm$^2$ energy range, spacecraft slewing 
capabilities of the order of 1 min and by assuming the X--ray afterglow 
photon 
fluence distribution measured by \swift/XRT, sensitive spectroscopy of tens of WHIM 
system per year could be done. In addition, by exploiting, e.g., resonant 
absorption lines, such instrumentation would allow the study of the galaxy ISM 
properties and their evolution with redshift.

\end{document}